\documentstyle[12pt,english]{article}
\newenvironment{eq}
{\[\begin{array}}{\end{array}\]{}}

\let\rvec=\vec        

   \def\({\Bigl(}
\def\){\Bigr)}    \def\|{\Big|}
\def\then{~\Rightarrow~}   \def\o{\circ}
    \def\x{\times}
   \def\ox{\otimes}
 
\def\pl{\oplus}

\def\PL{\displaystyle \bigoplus}

\def\mid{\big\bracevert}

\def\sub{\subseteq}
\def\subnoteq{\subset}

\def\supnoteq{\supset}
\def\and{\wedge}
\def\And{\bigwedge}
\def\AND{\displaystyle\bigwedge}
\def\od{\vee}

\def\rin{{\,\in\kern-.42em\in}}

\def\sign{{\,{\rm sign}\,}}

\def\rank{\,{\rm rank}}

\def\irrep{\,{\rm irrep}\,}

\def\tr{{\,{\rm tr }\,}}
\def\det{\,{\rm det }\,}
\def\id{\,{\rm id}}

\def\centr{\,{\rm centr}\,}

\def\sx{\rvec\x}

\def\A{{\,{\rm A\kern-.55emA}}}
\def\B{{\,{\rm I\kern-.2emB}}}
\def\C{{\,{\rm I\kern-.55emC}}}
\def\E{{\,{\rm I\kern-.2emE}}}
\def\G{{\,{\rm I\kern-.55emG}}}
\def\H{{\,{\rm I\kern-.2emH}}}
\def\I{{\,{\rm I\kern-.2emI}}}
\def\K{{\,{\rm I\kern-.2emK}}}
\def\L{{\,{\rm I\kern-.2emL}}}
\def\M{{\,{\rm I\kern-.16emM}}}
\def\N{{\,{\rm I\kern-.16emN}}}
\def\Q{{\,{\rm I\kern-.5emQ}}}
\def\R{{\,{\rm I\kern-.2emR}}}
\def\S{{\,{\rm I\kern-.42emS}}}
\def\T{{\,{\rm I\kern-.37emT}}}
\def\UU{{\,{\rm I\kern-.51emU}}}
\def\Z{{\,{\rm Z\kern-.32emZ}}}

\def\p{\partial}

\def\al{\alpha}  \def\be{\beta} \def\ga{\gamma}
\def\de{\delta}  \def\ep{\epsilon}  
   \def\vth{\vartheta} 
\def\ka{\kappa}   \def\la{\lambda}   \def\si{\sigma}
    
\def\phi{\varphi}  \def\Ga{\Gamma}  
    \def\La{\Lambda}

\def\vec#1{\underline{\bf vec}_{#1}}

\def\GL{{\bf GL}}  
\def\SL{{\bf SL}}
\def\U{{\bf U}} 
\def \UL{{\bf UL}} 
\def\O{{\bf O}}   
\def\SU{{\bf SU}} 
\def\SO{{\bf SO}}

 \def\D{{\bl D}}

\def\brack#1{\lbrack#1\rbrack}

\def\ro#1{{\rm #1}}
\def\bl#1{{\bf {#1}}}
\def\cl#1{{\cal #1}}

\def\ol#1{\overline{#1}}

\def\acom#1#2{\{#1,#2\}}

\def\map{\longrightarrow}

\def\lrmap{\leftrightarrow}

\def\dmap{\Big\downarrow}

\def\mape{\longmapsto}

\def\Diagr#1#2#3#4#5#6#7#8{\matrix{\noalign{\vskip5mm}
      &              &{\scriptstyle #5}&              &     \cr
      & #1           & \map           & #2           &     \cr
{\scriptstyle #8}   &\dmap         &    &\dmap  &{\scriptstyle#6} \cr
      & #4           & \map           & #3           &     \cr
      &              &{\scriptstyle#7}&              &     \cr
\noalign{\vskip5mm}             }}

\begin{document}

\begin{titlepage} 
\hfill MPI-PhT/98-35

\vskip3cm
\centerline{\bf THE EXTERNAL-INTERNAL GROUP} 
\centerline{\bf QUOTIENT STRUCTURE FOR THE STANDARD MODEL}
\centerline{\bf IN ANALOGY TO GENERAL RELATIVITY}
\vskip2cm
\centerline{
Heinrich Saller\footnote{\scriptsize 
saller@mppmu.mpg.de}  
}
\centerline{Max-Planck-Institut f\"ur Physik and Astrophysik}
\centerline{Werner-Heisenberg-Institut f\"ur Physik}
\centerline{M\"unchen}
\vskip25mm

\centerline{\bf Abstract}
\vskip5mm
In analogy to the class structure $\GL(\R^4)/\O(1,3)$ for general relativity
with  a local Lorentz group 
as stabilizer and a basic tetrad field for the  parametrization,  
a corresponding class structure $\GL(\C^2)/\U(2)$
is investigated for the standard model with a local hyperisospin group $\U(2)$.
The lepton, quark, Higgs and gauge fields,  used in the standard model,
cannot be basic in a coset interpretation, they may to be taken as
first order terms in a flat spacetime, particle oriented expansion of a 
basic field (as the analogue to the tetrad) and its
products.

\end{titlepage}

\newpage

\tableofcontents

\newpage

\advance\topmargin by -1.6cm

\section{The Coset Structure in  Relativity}

Usually, general relativity as the dynamics of a 
metric for a Lorentz manifold is
characterized with concepts from  differential geometry. 
To prepare a comparison of   relativity and the standard model from
a common  coset  point of view, I present in this section the 
well known \cite{UTIY}
Lorentz group class structure of relativity in a 
more algebraically oriented  language.

Special relativity distinguishes a Lorentz group 
$\O(1,3)$ 
with its causal order 
preserving orthochronous subgroup $\SO^+(1,3)$
as invariance group of a symmetric\footnote{\scriptsize
For a vector space $V$, the totally symmetric and antisymmetric 
tensor product subspaces are denoted 
with $V\od V$ and $V\and V$ resp. in 
the 2nd tensor power  $V\ox V$, correspondingly higher powers,
e.g. $V\od V\od V$ and $V\and V\and V$ in $V\ox V\ox V$, etc.} 
 pseudometric $g$  with signature $(1,3)$
on a real 4-di\-men\-sio\-nal vector space $\M\cong\R^4$  with spacetime 
translations (Min\-kow\-ski space)
\begin{eq}{l}
g:\M\od\M\map \R,~~\sign g=(1,3),~~g(v,w)=g(w,v)\cr
\O(1,3)\ni\La:\M\map\M\iff g=g\o (\La\od\La)
\end{eq}The inverse metric is used for the dual\footnote
{\scriptsize
$V^T$ denotes the algebraic dual with the linear forms for the vector space
$V$,
$f^T:W^T\map V^T$ is the dual (transposed) linear mapping for $f:V\map W$.
For finite dimensions, the 
linear mappings $\{f:V\map W\}$ are naturally isomorphic to the
tensor product
$W\ox V^T$.}
 energy-momentum space $\M^T$
\begin{eq}{l}
g^{-1}:\M^T\od\M^T\map\R,~~g^{-1}=g^{-1}\o(\La\od\La)^{-1T}
\end{eq}where the contragredient representation $\La^{-1T}$ acts on.

A Lorentz metric induces an isomorphism\footnote{\scriptsize
The sloppy notation  $g:\M\od\M\map\R$ and $g:\M\map\M^T$
with the same symbol $g\in\M^T\od\M^T$ should not lead to confusion.}
between
translations and energy-momenta
\begin{eq}{l}
g:\M\map\M^T,~~v\mape g(v,~),~~ g=g^T
\end{eq}It defines\footnote{\scriptsize
All diagrams are commutative.} 
a linear $g$-involution (Lorentz `conjugation') 
$f\stackrel g{\lrmap} f^g$ 
for all endomorphisms $f:\M\map\M$ of the translations
\begin{eq}{l}
\Diagr{\M}{\M^T}{\M^T}{\M}g{f^T}g{f^g},~~
\begin{array}{l}
f^g=g^{-1}\o f^T\o g\cr
f^{gg}=f\end{array}\cr
\hbox{for all }v,w\in\M:~~g(v,f(w))=g(f^g(v),w)\cr
\end{eq}The $g$-invariance Lorentz group 
is defined by $g$-unitarity\footnote{\scriptsize
Any involutive $g^{aa}=g\in G$ antiautomorphism $(gh)^a=h^ag^a$ of a group
$G$ defines the associate unitary subgroup $U(G,a)=\{g^a=g^{-1}\}$.
The inversion is the canonical antiautomorphism. In the quotient
$G/U(G,a)$ the unitary group is the stabilizer \cite{VIL}.} 
\begin{eq}{l}
\La\in\O(1,3)\iff\La^g=\La^{-1}\cr
\end{eq}The invariance 
Lorentz Lie algebra\footnote{\scriptsize
The  Lie algebra \cite{LIE13,RAIF} of a Lie group $G$ is denoted by $\log G$
which reminds also of log $\sim$ lag $\sim$ Lie algebra.}
is $g$-antisymmetric and, therefore, 
as a vector space isomorphic to the
antisymmetric square of the translations  
\begin{eq}{l}
l\in\log\O(1,3)\iff l^g=-l\cr
\R^{16}\cong\M\ox\M^T\supnoteq\log \O(1,3)\cong\M\and\M\cong\R^6
\end{eq}

There is a  manifold (symmetric space) $\GL(\R^4)/\O(1,3)$
of Lorentz groups in the
general linear group of a real 4-di\-men\-sio\-nal vector space
as illustrated 
by the 
 different invariance groups of the 
three metric matrices \cite{ALG9,FI} in one reference basis  of the translations

\begin{eq}{rcccc}
\hskip-4mm g\cong&
{\scriptsize\pmatrix{
g_{00}&g_{01}&g_{02}&g_{03}\cr 
g_{01}&g_{11}&g_{12}&g_{13}\cr 
g_{02}&g_{12}&g_{22}&g_{23}\cr 
g_{03}&g_{13}&g_{23}&g_{33}\cr 
}}, &
{\scriptsize\pmatrix{
1&0&0&0\cr 
0&-1&0&0\cr 
0&0&-1&0\cr 
0&0&0&-1\cr}},&
{\scriptsize\pmatrix{
0&0&0&1\cr 
0&-1&0&0\cr 
0&0&-1&0\cr 
1&0&0&0\cr}},&
{\scriptsize\pmatrix{
0&1&1&1\cr 
1&0&1&1\cr 
1&1&0&1\cr 
1&1&1&0\cr}}\cr
&g=g^T&\hbox{(time-space bases)}&\hbox{(light-space bases)}
&\hbox{(light bases)}\cr
&\sign g=(1,3)&\hbox{(Sylvester)}&\hbox{(Witt)}
&\hbox{(Finkelstein)}\cr
\end{eq}The manifold $\GL(\R^{1+s})/\O(1,s)$ with $s\ge1$ 
space dimensions can be visualized for
$s=1,2$ by all
possible $s$-di\-men\-sio\-nal 2-component hyperbola (hyperboloids) in $\R^{1+s}$. 

After the 
Stern-Gerlach experiment leading to 
the introduction of the spin operations with half integer 
$\SU(2)$-quantum numbers, also spacetime has to
come with a local `half-integer' Lorentz structure $\SL(\C^2)$.
The tetrad  field,
introduced by Weyl \cite{WEYL29} as the basic field for general relativity, 
maps a real 4-di\-men\-sio\-nal differentiable
spacetime manifold $\cl D$, parametrized
with four real coordinates $(x^\mu)_{\mu=0}^3\in\R^4$, into
the real 10-di\-men\-sio\-nal manifold of metrics.
It associates a $\GL(\R^4)/\O(1,3)$-class 
representative to 
each spacetime point  
\begin{eq}{l}
h:\cl D\map\GL(\R^4),~~ x\mape h(x)
\end{eq}It gives an isomorphism between the tangent space,
definable by the derivations $\ro{der}~\cl C(x)=\M(x)\cong\R^4$
of the differentiable functions at each spacetime point $x\in\cl D$,
 and one reference translation space  
$\M(0)\cong\R^4$ with metric $g(0)$
\begin{eq}{l}
h(x):\M(x)\map\M(0),~~
h^{-1T}(x):\M^T(x)\map\M^T(0)
\end{eq}Therewith
all multilinear\footnote{\scriptsize
$(\M\ox \M^T)(x)=\M(x)\ox\M^T(x)$ or 
$(\M\od \M)^T(x)=\M^T(x)\od\M^T(x)$ etc. are vector subspaces of the local
tensor algebra over $(\M\pl\M^T)(x)=\M(x)\pl\M^T(x)$.} 
 structures of $\M(0)$ and $\M(x)$
are bijectively related to each
other, e.g. the metric
and its  
invariance group
\begin{eq}{cc }
\Diagr{(\M\od\M)(x)}\R\R
{(\M\od\M)(0)}{g(x)}{\id_\R}{g(0)}{h(x)\od h(x)},&
\Diagr{\M(x)}{\M(x)}{\M(0)}{\M(0)}{\La(x)}{h(x)}{\La(0)}{h(x)}\cr
g(x)=g(0)\o(h\od h)(x),& \La(x)=h^{-1}(x)\o\La(0)\o h(x)\cr
\end{eq}

With dual $(\M(x),\M^T(x))$-bases,  e.g. 
$\{\p_\mu, dx^\mu\}$, one obtains as tensor components 
\begin{eq}{l}
h(x)\sim h_\mu^j(x)\sim h^T(x),~~h^{-1}(x)\sim h^\mu_j(x)
={\ep^{\mu\nu\rho\la}\ep_{jikl} h_\nu^i h_\rho^k h_\la^l\over 3!\det h}(x)
\sim h^{-1T}(x)\cr
g(0)\sim \eta_{jk},~~g^{-1}(0)\sim \eta^{jk}\cr 
g(x)\sim g_{\mu \nu}(x)=\eta_{jk}h_\mu^j h_\nu^k(x),
~~g^{-1}(x)\sim g^{\mu\nu}(x)\cr
\end{eq}With the Lorentz metric induced isomorphisms
between
tangent space and its dual, those relations
can be written in the form
\begin{eq}{l}
\Diagr{\M(x)}{\M^T(x)}{\M^T(0)}{\M(0)}
{g(x)}{h^{-1T}(x)}{g(0)}{h(x)},
\begin{array}{l}
g(0)\o h(x)\sim \eta_{jk}h^k_\mu(x)=h_{j\mu}(x)\cr
g^{-1}(0)\o h^{-1T}(x)\sim \eta^{jk}h^\mu_k(x)= 
h^{\mu j}(x)\end{array}
\end{eq}

Because of the invariance of the local metric under the local
Lorentz transformations
\begin{eq}{c}
g(x)=g(x)\o(\La\od\La)(x)=
g(0)\o(h\od h)(x)\o(\La\od\La)(x)
\end{eq}the tetrad field 
as coset representative is 
determined up to  local Lorentz transformations
\begin{eq}{l}
\La(x)\in\O(1,3)(x):~~ h(x)\mape h(x)\o\La(x)\end{eq}

This Lorentz gauge freedom of the tetrad  is made compatible 
with the translations (local derivations) by using an 
$\O(1,3)$-gauge field $\cl O(x)$ as a linear mapping from the translations into
the Lorentz Lie algebra
$\log\O(1,3)(0)$ of the reference space
\begin{eq}{l}
 \cl O(x):\M(x)\map(\M\ox\M^T)(0),~~\cl O(x)\sim O_ j^i{}_\mu(x)\cr
\end{eq}Because of the Lorentz invariance of the metric, 
a gauge field is $g(0)$-an\-ti\-sym\-me\-tric 
\begin{eq}{rl}
O(x):&\M(x)\map(\M\and\M)(0)\cr
&O(x)=g^{-1}(0)\o \cl O(x)\sim\eta^{ik}O_ k^i{}_\mu(x)=O_\mu^{ij}(x)=-O_\mu^{ji}(x)\cr
\end{eq}General relativity uses no
fundamental $\O(1,3)$-gauge field, but a `composite' one:
The local Lorentz freedom for the tetrad defines the 
tetrad induced gauge field $\cl O(x)=\cl O(h)(x)$ 
by using a  covariantly constant tetrad
\begin{eq}{rl}
Dh(x):&\M(x)\ox\M(x)\map\M(0)\cr
&Dh(x)=\p h(x)-h\o\Ga(x)- \cl O\o h(x)=0\cr
&D_\mu h^i_\nu(x)=
\p_\mu h^i_\nu(x)-h_\la^i\Ga^\la_{\mu\nu}(x)-O_\mu{}_ j^i h_\nu^j(x)=0\cr
\end{eq}with a  manifold connection $\Ga(x)$. 
A covariantly constant tetrad 
leads with $g(x)=g(0)\o (h\od h)(x)$ to a 
 covariantly constant metric
\begin{eq}{l}
Dg(x):\M(x)\ox(\M\od\M)(x)\map\R\cr
Dh(x)=0\then Dg(x)=0=
D_\mu g_{\nu\rho}(x)=
\p_\mu g_{\nu\rho}(x)
-\Ga_{\mu\nu}^\la g_{\la\rho}(x)
-\Ga_{\mu\rho}^\la g_{\nu\la}(x)\cr
\end{eq}If
the  $\log\GL(\R^4)$-valued connection  is assumed as
$g(x)$-symmetric (torsion free manifold),
it is expressable
by the tetrad and its derivative
\begin{eq}{l}
\hbox{if }\Ga^\la_{\mu\nu}(x)=\Ga^\la_{\nu\mu}(x)\then
\Ga_{\mu\nu}^\la(x)={g^{\la\rho}\over 2}(
\p_\mu g_{\nu\rho}
+\p_\nu g_{\mu\rho}
-\p_\rho g_{\mu\nu})(x)\cr
\end{eq}Therewith, the tetrad induced $\O(1,3)$-gauge field is determined 
\begin{eq}{rl}
\cl O(x)&=h^{-1}\o(\p h-h\o\Ga) (x)\cr
O_\mu^{ij}(x)&
=h^{\nu i}(\p_\mu h_\nu^j-h_\la^j\Ga^\la_{\mu\nu})(x)\cr
&={1\over2}h^{\la i}h^{\nu j}(
h_{\mu k}\p_{[\la}h_{\nu]}^k
+h_{\la k}\p_{[\mu}h_{\nu]}^k
-h_{\nu k}\p_{[\mu}h_{\la]}^k)(x)
\end{eq}

The tetrad induced $\O(1,3)$-curvature field $R(x)$ and $\cl R(x)$
relates the antisymmetric square of the tangent space 
and the local Lie algebra $\log\O(1,3)(x)\cong(\M\and\M)(x)$ to the 
antisymmetric square of the reference space and the reference Lie algebra
\begin{eq}{rl}
R(x):&
(\M\and\M)(x)
\map (\M\and\M)(0)\cr
&R(x)\sim R_{\mu\nu}^{ij}(x)=\p_{[\mu} O_{\nu]}^{ij}(x)
-O^{ik}_{[\mu}\eta_{kl}O^{lj}_{\nu]}(x)\cr
\cl R(x):&(\M\ox\M^T)(x)\map (\M\ox\M^T)(0)\cr
&\cl R(x)=g(0)\o R\o g^{-1}(x)\sim R_k^j{}_\mu^\la(x)
=\eta_{ki}R_{\mu\nu}^{ij}g^{\nu\la}(x)\cr
\end{eq}With the tetrad isomorphisms, it can be related to a transformation of the reference 
Lorentz Lie algebra
\begin{eq}{rl}
R\o (h\and h)^{-1}(x):&(\M\and\M)(0)\map (\M\and\M)(0)\cr
&R\o (h\and h)^{-1}(x)\sim R_{\mu\nu}^{ij}h^\mu_k h^\nu_l(x)\cr
\cl R\o (h\ox h^{-1})(x):&(\M\ox\M^T)(0)\map (\M\ox\M^T)(0)\cr
&\cl R\o (h\ox h^{-1})(x)\sim R_{\mu\nu}^{ij}h^\mu_k h^\nu_l(x)\cr
\end{eq}

The coupling of the curvature $\cl R$  to the tetrad $h\ox h^{-1}$
determines the familiar 2nd order derivative action 
\begin{eq}{l}
A(h,\p h)=\int \det h(x)~d^4 x ~\tr  \cl R\o (h\ox h^{-1})(x)\cr
\tr  \cl R\o (h\ox h^{-1})(x)
=  R_{\mu\nu}^{ij}h^\mu_i h^\nu_j(x)
=  R_k^j{}_\mu^\la h^\mu_j h_\la^k(x)=\tr  R\o (h\and h)^{-1}(x)
\cr
\end{eq}The integration over the manifold   
uses   the invariant volume element 
\begin{eq}{l}
{\AND^4}\M^T(x)\map{\AND^4}\M^T(0),~~
d^4 x\mape {\ep_{jikl}
h_\mu^j h^i_\nu h^k_\rho h^l_\la\over 4!}(x)
~dx^\mu\and dx^\nu\and
dx^\rho\and dx^\la 
\end{eq}

\section{The Linear Groups of  Relativity}

A Lorentz group $\O(1,3)$ is a semidirect product 
$\sx$ of a reflection group\footnote{\scriptsize
$\I(n)=\{z\in\C\mid z^n=1\}$ designates the $n$-th cyclotomic group.} 
$\I(2)=\{\pm 1\}$,
e.g. a time reflection,
and its special normal subgroup $\SO(1,3)$ which 
by itself  is the 
direct product $\x$ of the spacetime 
translations reflection group $\{\pm\bl 1_4\}\cong\I(2)$
and its orthochronous group $\SO^+(1,3)$
\begin{eq}{l}
\O(1,3)\cong\I(2)\sx\SO(1,3)\cong\I(2)\sx[\I(2)\x\SO^+(1,3)]\cr
\end{eq}The general linear group $g\in\GL(\R^4)$ contains 
via the modulus of the 4th root of the determinant
$|\sqrt[4]{\det g}|$
the abelian
dilatation group $\D(\bl 1_4)=\bl1_4\exp\R$ 
as a direct factor with the other factor $\UL(\R^4)$
(unimodular linear group) containing the elements with $|\det g|=1$
\begin{eq}{rl}
\GL(\R^4)&=\D(\bl1_4)\x\UL(\R^4)\cr
\UL(\R^4)&\cong\I(2)\sx\SL(\R^4)\cong
\I(2)\sx[\I(2)\x\SL_0(\R^4)]\cr
\end{eq}$\SO^+(1,3)=\SO_0(1,3)$ and
$\SL_0(\R^4)$ are the connection components of the group unit
in $\O(1,3)$ and $\UL(\R^4)$ resp. 
and the adjoint groups\footnote{\scriptsize
The adjoint group of a group $G$ consists of its classes 
$G/\centr G$ with respect to the
centrum.} of $\SO(1,3)$ and $\SL(\R^4)$ resp.

The tetrad manifold
is the product of the dilatation group and the quotient
of the connection components of the units 
\begin{eq}{l}
\GL(\R^4)/\O(1,3)\cong\D(1)\x\SL_0(\R^4)/\SO^+(1,3)\cr
\end{eq}The real 9-dimensional 
manifold $\SL_0(\R^4)/\SO^+(1,3)$ is the manifold of
nontrivial natural order structures $v\succeq 0$
on the translations $\M\cong\R^4$
as induced by the natural order of the scalars $\R$:
A natural translation order $\succeq$
has to be characterized by $\R$-multilinear forms, the even-linear
forms 
characterize  the pairs  $(\succeq,\preceq)$, 
consisting  of an order and its reverse.
Only the signature $(1,3)$-bilinear forms $g$ define  nontrivial order
pairs: $v\succeq 0$ or $v\preceq0$ $\iff$ $g(v,v)\ge0$.

The orbit $\{h^{-1}(x)\o\O(1,3)(0)\o h(x)\mid h(x)\in\GL(\R^4)\}$ 
of a reference Lorentz group
by inner automorphisms with $\GL(\R^4)$-operations 
does not fill the full group $\GL(\R^4)$ because of the nontrivial centralizer,
isomorphic to $\GL(\R)=\D(1)\x\I(2)$.

The equivalence classes $\irrep\SO^+(1,3)$ of the irreducible real 
finite dimensional representations of an
orthochronous Lorentz group 
with its simple rank 2 Lie algebra
are built by two fundamental representations, the
real 4-di\-men\-sio\-nal Minkowski representation $[1|1]$
(cyclic representation\footnote{\scriptsize
A cyclic representation generates by its tensor products 
all representations (up to equivalence).}),
selfdual with the symmetric signature $(1,3)$ Lorentz metric $g$,  and the
real 6-di\-men\-sio\-nal adjoint representation 
$[2|0]\pl[0|2]\cong[1|1]\and [1|1]$,
selfdual with two symmetric 
bilinear forms, the definite  metric $g\and g$
and the signature $(3,3)$-Killing metric\footnote{\scriptsize
The $\R^4$-volume element is a
symmetric bilinear form $\ep(4)\sim\ep^{ijkl}=\ep^{klij}$
with signature $(3,3)$ on $\R^4\and\R^4\cong\R^6$.}$\ep(4)$.
Correspondingly,
there are two types of 
real irreducible\footnote{\scriptsize
The representations $[2J_L|2J_R]\pl [2J_R|2J_L]$ are decomposable as 
complex representations.} 
 finite dimensional representations,
those with equal 
integer or half-integer 
`left' and `right' spin  numbers $J_L=J_R=J=0,{1\over2},1,\dots$ and
those with different `left' and `right' spin numbers $J_L\ne J_R$, but
integer sum
\begin{eq}{rll}
\irrep\SO^+(1,3)&=\{[2J|2J]\mid 2J=0,1,\dots\}&\cr
&\cup~\{[2J_L|2J_R]\pl [2J_R|2J_L]\mid\hskip-2mm& 
2J_{L,R}=0,1,\dots,~J_L\ne J_R,\cr
&&J_L+J_R=0,1,\dots\}
\end{eq}The dimensions for the representation spaces are
\begin{eq}{rl}
J_L=J_R=J:&\dim_\R[2J|2J]=(2J+1)^2\cr
J_L\ne J_R:&\dim_\R([2J_L|2J_R]\pl[2J_R|2J_L])=2(2J_L+1)(2J_R+1)\cr
\end{eq}All representations are selfdual, i.e. they have an 
$\SO^+(1,3)$-invariant 
bilinear form, symmetric as tensor product of the Lorentz metric.
 
The equivalence classes of the irreducible real 
finite dimensional representations \cite{FULHAR,HEL} of the
special  group 
$\SL_0(\R^4)$, locally isomorphic to 
$\SO(3,3)$, with a simple  rank 3 Lie algebra,
are built by three fundamental representations, the
real 4-di\-men\-sio\-nal cyclic representations $[1,0,0]$
and $[0,0,1]$, dual to each other, and the
real  6-di\-men\-sio\-nal  representation $[0,1,0]\cong[1,0,0]\and[1,0,0]$,
selfdual with the volume form $\ep(4)$
\begin{eq}{rl}
\irrep\SL_0(\R^4)&=\{[n_1,n_2,n_3]\mid n_{1,2,3}=0,1,\dots\}\cr
\dim_\R[n_1,n_2,n_3]&=
{(n_1+1)(n_3+1)(n_2+1)(n_1+n_2+2)(n_3+n_2+2)(n_1+n_3+n_2+3)\over 2!3!}\cr
\end{eq}The three natural numbers in $[n_1,n_2,n_3]$
are the linear combination coefficients 
of the dominant representation weight from the three fundamental
weights. The real 15-di\-men\-sio\-nal adjoint representation is
$[1,0,1]$.

The decomposition of the $\SL_0(\R^4)$-re\-pre\-sen\-ta\-tions into
$\SO^+(1,3)$-re\-pre\-sen\-ta\-tions is
given for the simplest cases, relevant in relativity 
\begin{eq}{l}\hskip-7mm
\begin{array}{|c||c|c|c|c|c|}\hline
\SL_0(\R^4)&\begin{array}{c}[1,0,0]\cr[0,0,1]\end{array}&[0,1,0]
&\begin{array}{c}[2,0,0]\cr[0,0,2]\end{array}&[0,2,0]&[1,0,1]\cr\hline
\hbox{dimension}&4&6={4\choose2}
&10={4+1\choose2}&20={6+1\choose2}-1&15=4^2-1={6\choose2}\cr\hline
\SO^+(1,3)&[1|1]&[2|0]\pl[0|2]
&[0|0]\pl[2|2]
&\begin{array}{l}[0|0]\pl[2|2]\pl\cr
[4|0]\pl[0|4]\end{array}
&\begin{array}{l}[0|0]\pl[2|2]\pl\cr
[2|0]\pl[0|2]\end{array}
\cr\hline
\end{array}
\end{eq}

The tangent space of the  tetrad (metric) manifold
is the  quotient of the corresponding  Lie algebras
\begin{eq}{l}
\log\GL(\R^4)/\log\O(1,3)\cong\M\od\M\cong \R^{10}
\end{eq}It
carries the irreducible representations
 $[2,0,0]$ of $\SL_0(\R^4)$. 
The curvature $R_{\mu\nu\ka\la}(x)=R^{ij}_{\mu\nu}h_{i\ka}h_{j\la}(x)$
with its familiar (anti)sym\-me\-try
properties 
as traceless element of $(\M\and\M)(x)\od (\M\and\M)(x)$
transforms with the 20-di\-men\-sio\-nal representation $[0,2,0]$,
the symmetric Ricci tensor $R_{\mu\la}(x)=R_{\mu\nu\ka\la}g^{\nu\ka}(x)$ 
with the 10-di\-men\-sio\-nal $[2,0,0]$.

In general, a representation $\psi$ of a group quotient $G/U$ 
will be defined as a mapping
from the classes $\psi:G/U\map V_U\ox V_G^T$ into the linear mappings
$\psi_{gU}:V_G\map V_U$ of two vector spaces with 
linear representations of the
groups involved, $G\map \GL(V_G)$ and $U\map \GL(V_U)$. If the vector spaces
are isomorphic $V_G\cong V_U\cong V$, the  mappings $\psi_{gU}\in\GL(V)$ are
assumed to be isomorphisms.

The tetrad $h(x),h^{-1}(x)\in\GL(\R^4)$ and the curvature 
$\cl R(x)\in\GL(\R^6)$
as representations of the quotient $\GL(\R^4)/\O(1,3)$ relate to each other 
vector spaces with the fundamental 
representations of the orthogonal and special  group.
In general, the $(n-1)$ fundamental $\SL_0(\R^n)$-representations
act on the $(n-1)$ Grassmann powers ${\AND^ N }\R^n$ for
$ N =1,\dots,n-1$. Therefore the reference 
Grassmann algebra\footnote{\scriptsize
$\And\M$ 
is isomorphic as vector space, not as associative algebra,
to the Clifford algebra over $\M$.}  
$\And\M(0)\cong\R^{16}$ over the translations with the 
powers ${\AND^ N }\M(0)\cong\R^{{4\choose  N }}$ 
as direct summands and the isomorphic local
partners $\And\M(x)$ are related 
to each other by the fields in relativity
\begin{eq}{c}
\begin{array}{|c||c|c|c|c|}\hline
 N &\hbox{Grassmann power }{\AND^  N }\M&\hbox{\bf field}&\SO^+(1,3)&\SL_0(\R^4)\cr\hline\hline
0&\R&\id_\R\sim 1&[0||0]&[0,0,0]\cr\hline
1&\M\cong\R^4&h(x)\sim h_\mu^j(x)&[1||1]&[1,0,0]\cr\hline
2&\M\and\M\cong\R^6&\cl R(x)\sim R^{ij}_{\mu\nu}(x)&[2||0]\pl [0||2]&[0,1,0]\cr\hline
3&\M\and\M\and\M \cong\R^4&h^{-1}(x)\sim h^\mu_j(x)&[1||1]&[0,0,1]\cr\hline
4&\R&\det h(x)&[0||0]&[0,0,0]\cr\hline
\end{array}\cr
\cr
\hbox{\it Lorentz and special linear  representation properties of the relativity fields}
\end{eq}The Grassmann degree $N$ 
is the $\D(1)$-grading,
by Weyl \cite{WEYLRZM}  called  `weight of a tensor density'.

\section{The Scales for  Relativity}

The rank of the symmetric space $\GL(\R^4)/\O(1,3)$
(tetrad or metric manifold)  
will be defined as the difference 
$4-2$ of the
ranks for the `nominator' and
`denominator' Lie algebra
\begin{eq}{l}
\rank_\R\GL(\R^4)/\O(1,3)=2, ~~\rank_\R \D(1)=1
\end{eq}The rank gives the number of invariants for the representations
of the manifold - one abelian invariant for $\D(1)$ and one simple
invariant for the quotient
$\SL_0(\R^4)/\SO^+(1,3)$. Those invariants can be used as overall normalization
and relative space-time normalization resp.  or as  
fundamental intrinsic length scale $\ell$ (Newton's constant)  and  
fundamental velocity scale ${\rm c}$ 
\begin{eq}{l}
g(x)\cong{\ell^2\over {\rm c}}
{\scriptsize\pmatrix{{1\over {\rm c}}&0\cr 0&-{\rm c}\bl1_3\cr}}
=h(\ell,{\rm c})
{\scriptsize\pmatrix{1&0\cr 0&-\bl1_3\cr}}h^T(\ell,{\rm c})\cr
h(\ell,{\rm c})={\scriptsize\pmatrix{{\ell\over {\rm c}}&0\cr 0&\ell\bl1_3\cr}}
~~~\hbox{ with }\ell,{\rm c}>0
\end{eq}The abelian invariant is given by the determinant 
of the tetrad  $h(\ell,{\rm c})$ or, in the
Lie algebra, by the trace, the simple invariant
arises from the `double trace' as familiar from the Killing form and  the
quadratic Casimir element for semisimple Lie algebras
\begin{eq}{l}
h(\ell,{\rm c})=\exp l(\ell,{\rm c}),~~\left\{\begin{array}{l}
\det h(\ell,{\rm c})=\exp\tr l(\ell,{\rm c})={\ell^4\over {\rm c}}\cr
\exp\sqrt{{ 4\tr l(\ell,{\rm c})\o l(\ell,{\rm c})
-(\tr l(\ell,{\rm c}))^2\over3}}={1\over {\rm c}}\end{array}\right.
\end{eq}

The flat spacetime expansion for general relativity  uses
the 10-di\-men\-sio\-nal  
tangent space of the tetrad manifold. It expands the $\GL(\R^4)$-tetrad 
 with its Lie algebra around a reference Lorentz
  group  $\O(1,3)$. A tetrad from the
unit connection component $\GL_0(\R^4)=\D(\bl 1_4)\x
\SL_0(\R^4)$
can be written with an exponent   
\begin{eq}{l}
h(x)=\exp l(x),~~l(x)\in\log\GL(\R^4)
\end{eq}Because of the local invariance, the Lie algebra element $l(x)$
is determined up to gauge translations $l(x)+\log\O(1,3)(x)$. 
The flat spacetime expansion 
is characterized by
\begin{eq}{l}
h(x)=\bl 1_4+ l(x)+\dots,~~h^j_\mu(x)=\de^k_\mu[\de_k^j+l^j_k(x)+\dots]
\end{eq}

\section{The Operation Groups of the Standard Model}
Before trying an 
interpretation with coset structures 
also for the standard model of the electroweak and
strong interactions, its relevant operational symmetries will be summarized.

The standard model implements the electroweak and strong interactions 
as gauge structures, relating the 
spacetime translations to  the
internal transformation groups
\begin{eq}{l}
\hbox{hypercharge: }\U(1),~~\hbox{isospin: }\SU(2),~~
\hbox{colour: }\SU(3)
\end{eq}In  the lepton, quark, Higgs and gauge fields,
the internal groups  meet 
with the external transformation groups\footnote{\scriptsize
The unspecified name `Lorentz group' is used for
the locally isomorphic real Lie groups
$\O(1,3)$, $\SO(1,3)$ (special), $\SO^+(1,3)$ (orthochronous)
and $\SL(\C^2)$ (covering).
The complex finite dimensional representations of the 
real dimension 6, rank 2 simple Lie algebra
$\log\SL(\C^2)$ are denoted with  2 natural numbers $[2J_L|2J_R]$
for the linear combination of its dominant weight from the 2 fundamental
weights for the Weyl representations.}
\begin{eq}{l}
\hbox{Lorentz group: }\SL(\C^2),~~\hbox{chirality: }\U(1) 
\end{eq}

The fundamental standard model fields  transform internally with
irreducible representations 
$\brack y$, $[2T]$ and $[ C_1, C_2]$ for hypercharge,
isospin  and colour group resp. and, externally, with
$\brack c$ and
$[2J_L|2J_R]$ for   chirality  and Lorentz group resp.
as given by the quantum numbers in the following table \cite{S981}
\begin{eq}{c}
\begin{array}{|c|c||c|c|c||c|c|}\hline
\hbox{\bf field}&\hbox{\bf symbol}&\U(1)&\SU(2)&\SU(3)&\U(1)&\SL(\C^2 )\cr
              &\bl\Psi&[y]&[2T]& [ C_1, C_2]&[c]&[2J_L|2J_R]\cr\hline\hline
\hbox{left lepton}&\bl l&-{1\over2}&[1]&[0,0]&{1\over2}&[1|0]\cr\hline
\hbox{right lepton}&\bl e&-1&[0]&[0,0]&{3\over2}&[0|1]\cr\hline
\hbox{left quark}&\bl q&{1\over6}&[1]&[1,0]&-{1\over2}&[1|0]\cr\hline
\hbox{right down quark}
&\bl d&-{1\over 3}&[0]&[1,0]&{1\over2}&[0|1]\cr\hline
\hbox{right up quark}
&\bl u&{2\over3}&[0]&[1,0]&-{3\over2}&[0|1]\cr\hline
\hbox{Higgs}&\bl H&-{1\over2}&[1]&[0,0]&1&[0|0]\cr\hline
\hbox{hypercharge gauge}&\bl A&0&[0]&[0,0]&0&[1|1]\cr\hline
\hbox{isospin gauge}&\bl B&0&[2]&[0,0]&0&[1|1]\cr\hline
\hbox{colour gauge}&\bl G&0&[0]&[1,1]&0&[1|1]\cr\hline
\end{array}\cr
\cr
\hbox{\it quantum numbers of the standard model fields}\cr
\end{eq}With respect to the Lorentz group, $[0|0]$
designates scalar fields, $[1|0]$ and $[0|1]$
are left and right handed Weyl spinor fields resp., $[1|1]$  vector fields.
The external and internal multiplicity
(singlet, doublet, triplet, quartet, octet, etc.)
 of the Lorentz-group,
isospin and colour  representations can be computed from the
natural numbers $2J_{L,R},2T, C_{1,2}$
\begin{eq}{rll}
\dim_\C[2J_L|2J_R]&=(2J_L+1)(2J_R+1),&2J_{L,R}=0,1,\dots\cr
\dim_\C[2T]&=2T+1,&2T=0,1,\dots\cr
\dim_\C[C_1,C_2]&={( C_1+1)( C_2+1)( C_1+
C_2+2)\over2},&C_{1,2}=0,1,\dots\cr
\end{eq}Fields and antifields have reflected quantum numbers
\begin{eq}{llrcr}
\bl \Psi&\hbox{with }&[y||2T; C_1, C_2]&\o&[c||2J_L|2J_R]\cr
\bl \Psi^*&\hbox{with }&[-y||2T; C_2, C_1]&\o&[-c||2J_R|2J_L]\cr
\end{eq}The chirality property $[c]$ will be discussed below in more detail.

The gauge  interaction of the fermion fields is effected by the 
local Lie algebra in\-va\-ri\-ants\footnote{\scriptsize
For a Lie algebra representation $\cl D:L\map V\ox V^T$ in the
endomorphism algebra of a vector space 
($L$ and $V$ finite dimensional) 
the  tensor  $\cl D\in V\ox V^T\ox L^T$ is the associated
invariant.}
 (current-gauge field products)
 \begin{eq}{l}  
g_1 \bl J(1)\bl A +g_2\bl J(2)\bl B + g_3\bl J(3)\bl G \cr
\begin{array}{ll}
\hbox{for }\U(1):&
\bl J(1)={1\over6}[\bl q^*\bl 1_6\bl q-2\bl d^*\bl 1_3\bl d 
-3\bl l^*\bl 1_2\bl l +4\bl u^*\bl 1_3\bl u -6\bl e^*\bl e]\cr 
\hbox{for }\SU(2):&
\bl J(2)={1\over2}[\bl q^*\rvec\tau\ox \bl1_3\bl q
+\bl l^*\rvec\tau\bl l]\cr
\hbox{for }\SU(3):&
\bl J(3)={1\over2}[\bl q^*\bl1_2\ox\rvec\la\bl q+\bl d^*\rvec\la\bl d 
 +\bl u^*\rvec\la\bl u]\end{array} 
\end{eq}involving as a basis e.g. the three Pauli and eight Gell-Mann matrices 
$\rvec\tau=(\tau^a)_{a=1}^3$ and $\rvec\la=(\la^c)_{c=1}^8$ resp.
The coupling constants $g^2_{1,2,3}>0$ are the normalizations of the
corresponding Lie algebras \cite{S981}.

At face value, the relevant group seems to be a product of five 
unrelated direct factors
\begin{eq}{l}
\underbrace{
\U(1)\x\SU(2)\x\SU(3)}_{\rm internal}\x\underbrace{
\U(1)\x\SL(\C^2)}_{\rm external}
\end{eq}A closer look, however,  suggests
a common origin for all those groups: 
The three internal factors are related to each other as well as the
two external ones and, highly interesting, there exists also an
internal-external correlation.

In general, a standard model field does not represent 
faithfully all operations.
If a group $G$ is represented, the faithfully represented group is
the quotient $G/N$, consisting of classes  with 
respect to the trivially represented 
invariant subgroup $N\sub G$.
To find those groups in the standard model, one has to consider
the four central correlations of its  operation group \cite{HUCK,S981}.

The two internal correlations connect hypercharge with both isospin and colour:
The colourless fields $\bl l$, $\bl e$, $\bl H$, $\bl A$ and $\bl B$
 show a (half)integer hypercharge-(half)integer
isospin correlation.
The isospin-less fields $\bl u$, $\bl d$ and $\bl G$ show an $\I(3)$
correlation.
Therefore the faithfully represented groups arise 
from the full unitary groups $\U(n)$ for $n=2,3$. $\U(n)$ 
is a product, not direct, of two normal subgroups 
with $\I(n)$ as discrete intersection\footnote{\scriptsize
The somewhat ambiguous notation ${G_1\x G_2\over H}$ 
denotes a common normal subgroup $H\sub G_1\cap G_2$
in contrast to e.g. $G_1\x G_2/H$.}.
Its quotient groups are the phase group 
$\U(\bl1_n)=\bl1_n\exp i\R$ and the
adjoint group $\SU(n)/\I(n)$
\begin{eq}{c}   
\left.\begin{array}{l}
\U(n)=\U(\bl1_n)\o\SU(n)\cr
\U(\bl1_n)\cap\SU(n)=\centr\SU(n)\cong\I(n)\end{array}\right\}\then
\U(n)\cong{\U(1)\x\SU(n)\over\I(n)}\cr
\end{eq}
\begin{eq}{c}   
\begin{array}{|c||c|c|}\hline
\hbox{normal subgroup}&\U(1)&\SU(n)\cr\hline
\hbox{quotient group}&\SU(n)/\I(n)&\U(1)\cr\hline\end{array}\cr
\cr
\hbox{\it internal operation groups from $\U(n)$, $n=2,3$}
\end{eq}Furthermore, the internal colour and isospin properties
of the left handed quark field $\bl q$ show that the internal faithfully
represented group, defined in $\U(6)$, 
is a product of three normal subgroups with an $\I(2)\x\I(3)
\cong\I(6)$
correlation 
\begin{eq}{l}
\U(2\x 3)=\U(\bl 1_6)\o[\SU(2)\ox\bl1_3\x\bl1_2\ox \SU(3)]\cr
\left.\begin{array}{l} \U(\bl 1_6)\cap[\SU(2)\ox\bl 1_3]\cong\I(2)\cr
\U(\bl 1_6)\cap[\bl 1_2 \ox\SU(3)]\cong\I(3)\end{array}\right\}
\then \U(2\x3) \cong{
\U(1)\x \SU(2)\x \SU(3)\over \I(2)\x\I(3)}\cr
\end{eq}
\begin{eq}{c}
\begin{array}{|c||c|c|c|}\hline
\hbox{normal subgroup}&\U(1)&\SU(2)&\SU(3)\cr\hline
\hbox{quotient group}&
\SO(3)\x\SU(3)/\I(3)&\U(3)&\U(2) \cr\hline\end{array}\cr
\cr
\begin{array}{|c||c|c|c|}\hline
\hbox{normal subgroup}&\U(2)&\U(3)&\SU(2)\x\SU(3)\cr\hline
\hbox{quotient group}&
\SU(3)/\I(3)&\SO(3)&\U(1)\cr\hline\end{array}\cr
\cr
\hbox{\it internal operation groups from $\U(2\x3)$}
\end{eq}

The external
correlation is seen in the fact that halfinteger spin $J_L+J_R$  comes with
halfinteger chirality number $c$ and integer $J_L+J_R$  with integer 
$c$. Therefore,
the faithfully represented external group is the 
unimodular group $\UL(2)=\{g\in\GL(\C^2)\mid|\det g|=1\}$
(phase Lorentz group).
Its quotient groups are the phase group 
(chirality group) and the orthochronous Lorentz group
as adjoint group
\begin{eq}{l}
\left.\begin{array}{l}
\UL(2)=\U(\bl1_2)\o\SL(\C^2)\cr
\U(\bl1_2)\cap\SL(\C^2)=\centr\SL(\C^2)\cong\I(2)\end{array}\right\}\then
\UL(2)\cong{\U(1)\x\SL(\C^2)\over\I(2)}\cr
\then \left\{
\begin{array}{rl}
\UL(2)/\SL(\C^2)&\cong\U(1)/\I(2)\cong\U(1)\cr
\UL(2)/\U(\bl1_2)&\cong\SL(\C^2)/\I(2)\cong\SO^+(1,3)\end{array}\right.
\end{eq}
\begin{eq}{c}
\begin{array}{|c||c|c|}\hline
\hbox{normal subgroup}&\U(1)&\SL(\C^2)\cr\hline
\hbox{quotient group}&\SO^+(1,3)&\U(1)\cr\hline\end{array}\cr
\cr
\hbox{\it external operation groups from $\UL(2)$}
\end{eq}

Before discussing the internal-external correlation,
the standard model fields will be arranged with respect to
the external and internal quotient groups of
$\UL(2)$ and $\U(2\x3)$ resp. they are representing  faithfully
\begin{eq}{c}
\begin{array}{|c||c|c||c|}\hline
&\UL(2)&\U(1)_{\rm ext}&\SO^+(1,3)\cr\hline\hline
\U(2)&\bl l&\bl H&\x\cr\hline
\U(1)_{\rm int}&\bl e&-&\x\cr\hline
\U(2\x3)&\bl q&-&\x\cr\hline
\U(3)&\bl d,\bl u&-&\x\cr\hline\hline
\SO(3)&\x&\x&\bl B\cr\hline
\{1\}&\x&\x&\bl A\cr\hline
\SU(3)/\I(3)&\x&\x&\bl G\cr\hline
\end{array}\cr\cr
\hbox{\it faithfully represented homogeneous groups in the standard model}
\end{eq}
  
Some entries are missing:
First of all, there are no coloured Lorentz 
scalar  fields, analogous to the Higgs isodoublet.

Secondly: 
A field of the standard model has
nontrivial hypercharge if, and only if, it has
nontrivial chirality.
The chirality $\U(1)_{\rm ext}$ number $c$ is determined from the 
Yukawa interaction 
\begin{eq}{l}
(\mu_e\bl e^*\bl l+\mu_u\bl q^*\bl u+\mu_d\bl d^*\bl q)\bl H
+\hbox{h.c.}~~\hbox{ with Yukawa couplings }\mu_{e,u,d}\in\R
\end{eq}With an integer $c_{\bl H}$ for the Higgs field,
the chiral numbers for  the quark fields $\bl q,\bl d,\bl u$
and  for the lepton fields $\bl l,\bl e$
are given up to integers  $z_q$ and $z_l$ resp.
\begin{eq}{c}
\begin{array}{|c||c||c|c||c|}\hline
&\U(1)_{\rm int}&\U(1)_{\rm ext}&\U(1)_{\rm ext}\hbox{ with}&\U(1)_{\rm ferm}\cr
&y&c&c_{\bl H}=1,~z_{q,l}=0&f=-c-2y\cr\hline\hline
\bl l&-{1\over2}&{1\over2}+z_l&{1\over2}&{1\over2}\cr\hline 
\bl e&-1&{1\over2}+z_l+c_{\bl H}&{3\over2}&{1\over2}\cr\hline 
\bl q&{1\over6}&-{1\over2}+z_q&-{1\over2}&{1\over6}\cr\hline 
\bl d&-{1\over3}&-{1\over2}+z_q+c_{\bl H}&{1\over2}&{1\over6}\cr\hline 
\bl u&{2\over3}&-{1\over2}+z_q-c_{\bl H}&-{3\over2}&{1\over6}\cr\hline 
\bl H&-{1\over2}&c_{\bl H}&1&0\cr\hline 
{\bf A,B,G}&0&0&0&0\cr\hline 
\end{array}\cr
\cr
\hbox{\it hypercharge, chirality and fermion numbers for the standard model
fields} 
\end{eq}The choice of the three integers $c_{\bl H},z_l,z_q$ is not obvious.
$z_l$ and $z_q$ will be determined 
by opposite chirality and hypercharge for the lepton isodoublet 
field $\bl l$ and opposite chirality and
threefold hypercharge for the quark isodoublet field $\bl q$  
\begin{eq}{l}
c_{\bl l}=-y_{\bl l},~~
c_{\bl q}=-3y_{\bl q}\then z_l,z_q=0
\end{eq}The chirality $c_{\bl H}$ for the Higgs field
is determined in such a way that the hypercharge-chirality combination 
(fermion number) 
$ f=-c-2c_{\bl H}y$, trivial for the Higgs field, gives a ratio $1:3$
for quark and lepton fields
\begin{eq}{l}
f_{\bl l}=3 f_{\bl q}\iff 
c_{\bl l}+2c_{\bl H} y_{\bl l}=3(c_{\bl q}+2c_{\bl H} y_{\bl q})\then c_{\bl H}=1
\end{eq}Those conditions will be discussed
in section 4 and 6. 

Both $\U(1)$'s,
chirality and hypercharge, have to be represented in the only one phase group
of a field. 
The combination of chirality and hypercharge
with trivial value for the Higgs field  defines 
a fermion number group $\U(1)$ which correlates external and
internal $\U(1)$ 
\begin{eq}{l}
\left.\begin{array}{rl}
\U(1)_{\rm ext}&\subnoteq{\bf UL}(2)\cr
\U(1)_{\rm int}&\subnoteq\U(2\x3)\end{array}\right\},~~
\U(1)_{\rm ferm}\cong{\U(1)_{\rm ext}\x\U(1)_{\rm int}\over\U(1)}\cr
f=-c-2y=\left\{\begin{array}{rl}
{1\over2}&\hbox{for lepton fields }\bl l,~\bl e\cr
{1\over6}&\hbox{for quark fields }\bl q,~\bl d,~\bl u\cr
0&\hbox{for boson fields }\bl H,~\bl A,~\bl B,~\bl G\cr
\end{array}\right.
\end{eq}

Summarizing the operation groups of the standard model:
The external-internal homogeneous symmetry group,
faithfully represented with the standard model fields,
is a  product of five normal subgroups with a fourfold central correlation
\begin{eq}{l}
{ \U(2\x 3)\x\UL(2)\over\U(1)}
\cong{\U(1)\x\SU(2)\x\SU(3)\x\U(1)\x\SL(\C^2)\over
\I(2)\x\I(3)\x\U(1)\x\I(2)}
\end{eq}

\section{Symmetries for Particles}
One has to make a clear distinction between the operation group (symmetry)
for fields and the operation group (symmetry) for particles \cite{WIG}:
Going from the standard model fields for the description of the dynamics to the
in- and out-fields for the description of  particles, the 
homogeneous  real 18-dimensional Lie group
${ \U(2\x 3)\x \UL(2)\over\U(1)}$ with both external and internal operations 
is dramatically reduced. With colour confinement and ground state frozen
electroweak symmetries there remains from the 12-dimensional $\U(2\x3)$ only a
1-dimensional abelian $\U(1)$-symmetry, 
faithfully represented by  particles with nontrivial electromagnetic charge
or fermion number, e.g. by the electron or the neutron. 
The establishment of a laboratory distinguishes a 
reference rest system and reduces 
the 6-dimensional external Lorentz group operations $\SL(\C^2)$
for fields in the case of
 massive halfinteger and integer spin particles  
to a faithfully represented  3-dimensional  group $\SU(2)$ and
$\SU(2)/\I(2)\cong \SO(3)$ resp. Massless  
particles represent faithfully only a 1-dimensional polarization subgroup 
$\SO(2)\cong\U(1)\subnoteq\SU(2)$, which - possibly reflecting 
the external-internal $\U(1)$-correlation - are all chargeless,
e.g. the photon and the neutrinos 
\begin{eq}{c}\hskip-6mm
\begin{array}{|c||c|c|c|c|c|}\hline
&&\D(1)&\U(1)\subnoteq\SU(2)&\U(1)&\U(1)\cr 
 \hbox{\bf particle}&\hbox{\bf symbol} &\hbox{mass}&\hbox{spin}&\hbox{polarization}&\hbox{el.mgn.}\cr
& &&\hbox{direction}
&\hbox{(helicity)}&\hbox{charge}\cr
              \hline\hline
\hbox{massive electron}&\ro e^\mp&m_{\ro e}&+{1\over2},-{1\over2}&-&\mp 1\cr\hline
\hbox{electron neutrino}&\nu_{\ro e},\ol\nu_{\ro e}&0&-&\pm 1&0\cr\hline
\hbox{charged weak boson}&\ro W^\pm&m_{\ro W}&+1,0,-1&-&\pm 1\cr\hline
\hbox{neutral weak boson}&\ro Z&m_{\ro Z}&+1,0,-1&-&0\cr\hline
\hbox{photon}&\ga&0&-&\pm 1&0\cr\hline
\end{array}\cr\cr
\hbox{\it particles from standard fields}
\end{eq}

\section{The Coset Structure in the Standard Model}

After the coset formulation for relativity in sections 1,2,3 and the
exposition of the standard model operation groups in section 4, I come to the main
purpose of this paper.

An attempt to characterize the standard model for the electroweak and
strong interactions with coset structures and symmetric spaces in analogy to
relativity  encounters characteristic differences:
Relativity is a real theory with orthogonal groups and bilinear forms
 (metrics) whereas the standard model and quantum theory come in a complex
formulation with unitary groups and sesquilinear forms 
(scalar products, probability amplitudes).   
The local operation Lorentz group $\O(1,3)$ for relativity has no true 
normal Lie subgroup whereas the internal standard model operation group 
$\U(2\x3)$ has
the normal Lie subgroups $\U(1)$ (hypercharge), 
$\SU(2)$ (isospin) and $\SU(3)$ (colour). The main
apparent obstacle for a symmetric space interpretation for the standard model
is the colour group $\SU(3)$: It prevents a naive embedding of the
internal group $\U(2\x3)$ as subgroup of the external phase Lorentz group  $\UL(2)$
- as compared  to
the tetrad manifold quotient structure $\GL(\R^4)/\O(1,3)$.
Therefore, Weinberg's `Model of Leptons' \cite{WEIN} is considered first: 
There, the colourless group ${\U(2)\x\UL(2)\over\U(1)}$ 
with    hyperisospin and phase Lorentz group  is  represented by
the lepton fields
$\bl l$, $\bl e$, the 
hypercharge and isospin gauge fields $\bl A,\bl B$ and the Higgs field
$\bl H$. 

A group
$\U(2)$ (hyperisospin) is the invariance group of a
definite   scalar product $d$ for a complex 2-di\-men\-sio\-nal
vector space $\UU\cong\C^2$ 
\begin{eq}{l}
d:\UU\x\UU\map \C,~~d(v,v)>0\iff v\ne 0,~~d(v,w)=\ol{d(w,v)}\cr
\U(2)\ni u:\UU\map\UU\iff d=d\o(u\x u)\cr
\end{eq}A scalar product $d$ for quantum theory
is the analogue to a signature $(1,3)$ metric $g$ of the
real translation vector space $\M\cong\R^4$ in relativity 
with $\O(1,3)$-invariance (section 1).

A scalar product
defines a conjugation $f\stackrel d{\lrmap} f^*$ 
 for all linear mappings  $f:\UU\map\UU$ 
\begin{eq}{l}
\hbox{for all }v,w\in\UU: d(v,f(w))=d(f^*(v),w),~~f^{**}=f
\end{eq}with $u\in\U(2)\iff u^*=u^{-1}$ and $l\in\log\U(2)\iff l=-l^*$.

Antilinear structures like a sesquilinear complex scalar product
$d$ are more complicated than linear ones.
In general for a complex linear space $\UU\cong\C^n$, one has 
to consider the
complex quartet\footnote{\scriptsize
The complex quartet structure leads also to the  fourfold
concept `particle creation,
 particle annihilation,  antiparticle creation and antiparticle annihilation'.}
 of associated vector spaces 
$\UU,\UU^T,\UU^*,\UU^{*T}\cong\C^n$,
consisting   of
space, dual space, antispace and dual antispace resp. \cite{ALG13,HAFT},
 to
take care of the conjugations in a basic independent
form. The canonical $\C$-conjugation
defines canonical antilinear isomorphisms between antispaces
$\UU\cong\UU^*$ and
$\UU^*\cong\UU^{*T}$. 
With an additional vector space  conjugation, i.e. an antilinear isomorphisms 
 between duals, $d:\UU\map\UU^T$, $v\mape d(v,~)$, 
 one obtains linear isomorphisms $\UU\cong\UU^{*T}$ and $\UU^T\cong \UU^*$.

There is a real 4-di\-men\-sio\-nal manifold  (symmetric space) 
$\GL(\C^2)/\U(2)$  of
positive  unitary
groups in the general linear group,
considered as real 8-di\-men\-sio\-nal Lie group.
With a reference basis, this manifold is pa\-ra\-me\-tri\-zab\-le by 
all positive $2\x2$-matrices  for the scalar products
\begin{eq}{l}
 d\cong
 {\scriptsize\pmatrix{
 d_0+ d_3& d_1-i d_2\cr
 d_1+i d_2& d_0- d_3\cr}}\succ0 \iff\left\{\begin{array}{l}
 d= d^*\hbox{ and }\tr d,~\det d>0\cr
\hbox{i.e. } d_j\in\R\hbox{ and } d_0,~ d^2=d_0^2-\rvec d^2>0\end{array}
\right.
\end{eq}In analogy to  $\al=\ep(\al)\al$
for a positive number $\al>0$,
the positivity of the matrix $d$ is expressable  with 
its signature $\ep(d)=\ep(d_0)\vth(d^2)$
\begin{eq}{l}
d\succ0\iff d\ne0\hbox{ and }d=\ep(d_0)\vth(d^2) d 
\end{eq}

Besides the analogies, there are important
differences between the 
real-orthogonal quotient structure of relativity and 
the complex-compact one proposed  for the standard model:
In contrast to the different dimensions of 
the spacetime and tetrad manifold in  relativity for $\M\cong\R^4$
\begin{eq}{l}
4=1+s=\dim_\R\cl D<\dim_\R\GL(\R^{1+s})/\O(1,s)={2+s\choose2}=10
\end{eq}one has coinciding dimensions for 
4-dimensional spacetime and
scalar product manifold for $\UU\cong\C^2$
\begin{eq}{l}
\dim_\R\cl D=\dim_\R\GL(\C^n)/\U(n)=n^2=4
\end{eq}Consequently, the symmetric space $\D(2)$
can be used \cite{S97} as a model for the spacetime manifold 
\begin{eq}{l}
\cl D=\D(2)=\GL(\C^2)/\U(2)\cong\exp\R^4
\end{eq}With this interpretation, spacetime arises as the manifold of 
compact operations $\U(2)$
in general linear operations $\GL(\C^2)$.

The full linear group $\GL(\C^2)$ is the direct product of its
dilatation group
$\D(\bl1_2)=\bl1_2\exp\R$ and its unimodular group $\UL(2)$
\begin{eq}{rl}
\GL(\C^2)&=\D(\bl1_2)\x\UL(2)\cr
\D(2)&=\GL(\C^2)/\U(2)\cong\D(1)\x{\bf SD}(2)\cr
\end{eq}The spacetime manifold $\D(2)$  involves as direct nonabelian factor the 
real 3-di\-men\-sio\-nal boost manifold 
\begin{eq}{rl}
{\bf SD}(2)=\UL(2)/\U(2)\cong\SL(\C^2)/\SU(2)
\cong\SO^+(1,3)/\SO(3)
\end{eq}The tangent spaces of
the homogeneous space as the quotient of the correponding Lie algebras
\begin{eq}{l}
\log\GL(\C^2)/\log\U(2)\cong\M\cong \R^4
\end{eq}can be taken for the  Minkowski
translations  carrying the irreducible 
$\SL(\C^2)$-re\-pre\-sen\-ta\-tions $[1|1]$ of the
adjoint group $\GL(\C^2)/\GL(\C)\cong\SO^+(1,3)$. 
The Cartan representation of the 
spacetime translations $\M$ by $\U(2)$-hermitian
complex $2\x 2$-matrices $x=x^*$ 
shows the local $\U(2)$-structure
\begin{eq}{l}
\M\cong \D(2)=\exp\M,~~\R^4\cong\exp\R^4\cr
x=x^*={\scriptsize\pmatrix{
x_0+x_3&x_1-ix_2\cr x_1+ix_2&x_0-x_3\cr}}\cr
d(x)=\exp x=(\cosh|\rvec x|+{\rvec\si\rvec x\over|\rvec x|}
\sinh|\rvec x|)\exp x_0={\scriptsize\pmatrix{
 d_0(x)+ d_3(x)& d_1(x)-i d_2(x)\cr
 d_1(x)+i d_2(x)& d_0(x)- d_3(x)\cr}}
\end{eq}

In the special manifold factors $\SL_0(\R^4)/\SO^+(1,3)$
(manifold of natural orders) and
$\SL(\C^2)/\SU(2)$
(manifold of conjugations), the orthogonal stability group $\SO^+(1,3)$ has a
signature $(1,3)$ invariant Lorentz form $g$ 
on the translations $\M\cong\R^4$ whereas the unitary group $\SU(2)$ has, in
addition to an invariant scalar product $d$ on $\U\cong\C^2$, an invariant
antisymmetric bilinear form $\ep(v,w)=-\ep(w,v)$ (`spinor metric'). The
$\C^2$-volume form $\ep$ is invariant also with respect to
$\SL(\C^2)$, it leads to the bilinear  symmetric
orthochronous $\SO^+(1,3)$-forms 
$g\cong\ep\ox\ep^{-1}$.
No $\SL_0(\R^4)$-invariant bilinear form exists on 
the translations $\M$.

\section{Spacetime as Basic Field Quantization}

In analogy to the  relativity tetrad $h$ as basic representation of 
the real 10-di\-men\-sio\-nal metric manifold $\GL(\R^4)/\O(1,3)$, a
basic field  $\psi$ is introduced as fundamental representation for the 
real 4-di\-men\-sio\-nal manifold $\GL(\C^2)/\U(2)$ of scalar products. 
It associates to each point of the real 4-di\-men\-sio\-nal spacetime  
$\cl D=\D(2)$, 
parametrizable with $d(x)=\exp x$ for $x\in\R^4$,  a  class representative
\begin{eq}{l}
\psi:\D(2)\map\GL(\C^2),~~d(x)\cong x\mape\psi(x)
\end{eq}With the basic field $\psi$, a complex vector space $\UU(x)\cong\C^2$ 
at each spacetime point can be related to 
a reference space. $\psi^*(x)$ gives an isomorphism
between the reference antispace $\UU^*(0)$ and the antispace
 $\UU^*(x)$
\begin{eq}{l}
\psi(x):\UU(x)\map\UU(0),~~\psi^{*-1}(x):\UU^*(x)\map\UU^*(0)
\end{eq}with the  scalar products
\begin{eq}{l}
\Diagr{(\UU\x\UU)(x)}\C\C{(\UU\x\UU)(0)}{ d(x)}{\id_\C}
{ d(0)}{(\psi\x\psi)(x)},~~ d(x)= d(0)\o(\psi\x\psi)(x)
\end{eq}Bases are given with  $\al,A=1,2$ 
\begin{eq}{l}
\psi(x)\sim\psi^\al_A(x)\sim\psi^T(x),~~
 \psi^{*-1}(x)\sim\psi^{* \dot A}_\be(x)=
\de_{\al\be}\psi^{* \al}_A(x)\de^{A\dot A}\sim\psi^{*-1T}(x)\cr
d(0)\sim \de_{\al\be}\cr
 d(x)\sim  d_{AB}(x)=
\de_{\be\al}\psi^{*\al}_A\psi^\be_B(x)
\cong d_B^{\dot A}(x)=d_{AB}(x)\de^{A\dot A}
=\psi^{* \dot A}_\be\psi^\be_B(x)\cr
\end{eq}The basic fields
$\psi$ and $\psi^*$ transform under the  
two conjugated fundamental complex 2-di\-men\-sio\-nal 
$\UL(2)$-re\-pre\-sen\-ta\-tions (left and
right handed Weyl spinors),
as usual denoted with undotted and dotted indices.

The  $\GL(\C^2)/\U(2)$-analogue 
to the flat spacetime expansion in general relativity 
$\GL(\R^4)/\O(1,3)$ with the
tetrad expansion $h=\bl 1_4+\dots$ around a reference $\O(1,3)$ 
requires an expansion of
the external group $\UL(2)$ around 
a compact local reference group $\U(2)$. Such an
expansion in the standard model is performed by the transition 
from the operation group representing fields for the dynamics to the 
tangent particle fields 
(in- and out fields) involving the dramatic
symmetry reduction mentioned above
and requires the definition of a ground state
and a reference system (spontaneous symmetry breakdown). 

By an
expansion of the coset representative $\psi$ in  
flat spacetime $\M\cong\R^4$  
with  the standard model lepton  fermion field $\bl l$   
\begin{eq}{rlll}
\psi(x)&=\bl l(x)+\dots,~~~&\psi_A^\al(x)&=\bl l_ A^\al(x)+\dots\cr
\psi^*(x)&=\bl l^*(-x)+\dots,
&\psi^{*\dot A}_\al(x)&=\bl l^{*\dot A}_\al(-x)+\dots
\end{eq}the spacetime defining scalar product can be related to
the anticommutator quantization condition\footnote{\scriptsize
For the left-handed part of the massive lepton particle field one has the
anticommutator
\begin{eq}{l}
\acom{\bl l^*(0)}{\bl l(x)}=\int{d^4q\over8\pi^2}q\ep(q_0)\de(q^2-M^2)
\exp iqx=x\ep(x_0)[\de'(x^2)-{M^2\over4}\de(x^2)+{M^4\over16}\vth(x^2)+\dots]
\end{eq}} 
\begin{eq}{l}
\begin{array}{rlll}
\log d(x)&= \acom    {\psi^*(x)}{\psi(x)}&= 
\acom    {\bl l^*(-x)}{\bl l(x)}+\dots 
&= x\ep(x_0)\de'(x^2)+\dots \cr
\log d^{\dot A}_B(x)&= \acom    {\psi^{*\dot A}_\be(x)}{\psi^\be_B(x)}&= 
\acom    {\bl l^{*\dot A}_\al(-x)}{\bl l^\al_B(x)}+\dots 
&= x^{\dot A}_B\ep(x_0)\de'(x^2)+\dots\end{array}  \cr
\hbox{ with }
x=x^*\sim x^{\dot A}_B=(\si^j)_B^ {\dot A} x_j=
{\scriptsize\pmatrix{
x_0+x_3&x_1-ix_2\cr x_1+ix_2&x_0-x_3\cr}}
\end{eq}With the canonically quantized flat space standard model fields alone a coset
interpretation breaks down at this point.
A  quantization involving lightcone supported distributions does not allow  
an interpretation 
as a spacetime dependent scalar product  
$d(x)$. Additional nonparticle contributions \cite{S96,S97}
can lead to an expansion for the basic field quantization
without lightcone supported distribution
\begin{eq}{l}
\log d(x)= \acom    {\psi^*(x)}{\psi(x)}=  x\ep(x_0)\vth(x^2)+\dots \cr
\end{eq}The parametrization of the spacetime manifold $\D(2)$ is effected by
the quantization of the basic field $\psi$.

\section{The Scales for the Standard Model}

The representations of the scalar product  manifold 
$\D(2)$ with 
real rank 2 as model for spacetime are
characterized by two real invariants - an abelian dilatation 
invariant $M$  and
a simple `boost'-invariant $m$
\begin{eq}{l}
\rank_\R\GL(\C^2)/\U(2)=\rank_\R\D(1)+\rank_\R\SO^+(1,3)/\SO(3)=2\cr
d(x)\cong \exp M
{\scriptsize\pmatrix{\exp
m&0\cr0&\exp(-m)\cr}}
=\psi(M,m){\scriptsize\pmatrix{1&0\cr0&1\cr}}\psi^*(M,m)\cr
\end{eq}The two invariants, given in the Lie algebra structure
by the abelian trace and the simple `double trace'
\begin{eq}{rlrl}
\psi(M,m)=\hskip-2mm&\exp {M\over2}{\scriptsize\pmatrix{\exp
{m\over2}&0\cr0&\exp(-{m\over2})\cr}},&
\psi(m)=\hskip-2mm&{\psi(M,m)\over|\sqrt{\det \psi(M,m)}|}={\scriptsize\pmatrix{\exp
{m\over2}&0\cr0&\exp(-{m\over2})\cr}}\cr
M=\hskip-2mm&\tr\log \psi(M,m),&
m^2=\hskip-2mm&2\tr\log \psi(m)\o\log \psi(m)
\end{eq}can be used \cite{S97} as
fundamental mass scale $M$ and fundamental interaction range ${1\over m}$
in the representations of the spacetime manifold 
$\D(2)\cong\D(1)\x{\bf SD}(2)$ by quantum fields.

\section{Hyperisopin Gauge Fields}

The curvature in relativity $\cl R:(\M\ox\M^T)(x)\map
(\M\ox\M^T)(0)$ relates 
to each other the Lorentz Lie algebras acting on the tangent spaces. 
The analogue for the standard model considers 
the tensor product 
 $\UU\ox\UU^*\cong\C^4$  for  a scalar product space $\UU$ 
with the represented group $\U(2)$
and its Lie algebra.
The two real 4-di\-men\-sio\-nal subspaces $\{f=\pm f^*\mid f\in\UU\ox\UU^*\}$
of the endomorphisms $\UU\ox\UU^*\cong\R^4\pl i\R^4$ are both stable
under  the action of $\U(2)$. The product representation of $\U(2)$
decomposes into a 3-dimensional representation, faithful for 
the adjoint group $\SO(3)\cong\U(2)/\U(1)$, and a 1-dimensional
trivial one 
\begin{eq}{rl}
u\in\U(2):&u\ox u:\UU\ox\UU^*\map \UU\ox\UU^*\cr
&u\ox u\cong \id_\C\pl O_3(u)\in\{1\}\pl\SO(3)\cr
\end{eq}

$\U(2)$-gauge fields $\cl G$ 
associate to each spacetime point 
an isomorphism to a reference tensor space
\begin{eq}{rl}
\cl G(x):&(\UU\ox\UU^*)(x)\map(\UU\ox\UU^*)(0)\cr
&\begin{array}{rl}
\cl G(x)=(\psi\ox\psi^*)(x)&=\cl A(x)\pl\cl B(x)\cr
\cl G^{ \dot B\al}_{A\be}(x)=\psi^\al_A\psi^{*\dot B}_\be(x)&=
(\si^j)^{\dot B}_A [\de^\al_\be \cl A_j(x)+\rvec \tau^\al_\be 
\rvec {\cl B}_j(x)]\cr
\end{array}\cr
\end{eq}Therewith, the  manifold of 
$(1\pl 3)$-decomposable 4-di\-men\-sio\-nal
isospin $\SO(3)$-re\-pre\-sen\-ta\-tions  
on the tensor product  is considered 
in the orthochronous Lorentz group $\SO^+(1,3)\cong\UL(2)/\U(1)$.

The hyperisospin $\U(2)$-gauge fields  of the standard model
might be taken as one term in the particle oriented flat spacetime approximation
\begin{eq}{rl}
\cl A_j(x)=
{1\over4}\psi^\al_A\de_\al^\be (\ol\si_j)_{\dot B}^A\psi^{*\dot B}_\be(x)
&=\bl A_j(x)+\dots\cr
\rvec{\cl B}_j(x)=
{1\over4}\psi^\al_A\rvec\tau _\al^\be (\ol\si_j)_{\dot B}^A\psi^{*\dot B}_\be(x)
&=\rvec{\bl B}_j(x)+\dots 
\end{eq}

In general, the standard model fields seem to be 
the particle  related and ground state respecting  contributions
 in a flat spacetime expansion 
for  the  more basic fields $\psi,\psi^*$ which pa\-ra\-me\-tri\-ze
the $\U(2)$-operations in $\UL(2)\subnoteq\GL(\C^2)$ 
acting on  the  tensor powers of the  
vector spaces $\UU,\UU^*$. This is  
done for the basic space
$\UU$  with faithful  hyperisospin $\U(2)$ action by
the standard lepton field
$\psi=\bl l+\dots$ and for
the tensor space  $\UU\ox\UU^*$ with adjoint 
isospin group $\U(2)/\U(1)$-action by the standard
hypercharge and isospin gauge fields $\psi\ox\psi^*=\bl A\pl\bl B+\dots$.  

\section{The Grassmann Algebra for Spacetime}

The local Grassmann algebra 
$\And\M\cong\R^{16}$ over the translations
at each point of the spacetime manifold 
in relativity has as analogue the
local Grassmann algebra 
over $\UU\pl\UU^*\cong\C^4$ for the standard model. 
In contrast to the translations $\M\cong\R^4$,
the vector space $\UU\cong\C^2$ does not arise as a tangent space.
The totally antisymmetric  tensor powers 
${\AND^ N }(\UU\pl\UU^*)$ with Grassmann degree $  N =0,1,2,3,4$ 
carry  all fundamental representations
of hyperisospin $\U(2)$ and its quotient groups $\U(1)$ and $\SO(3)$. Their
direct sum constitutes the complex 
Grassmann (exterior)
 algebra \cite{S922,S924} $\And(\UU\pl \UU^*)={\bf GRASS}\cong\C^{16}$
\begin{eq}{c}
\begin{array}{|c||c|c|c|}\hline
 N &\begin{array}{c}\hbox{subspaces of}\cr
{\AND^ N }(\UU\pl\UU^*)\cong\C^{{4\choose  N  }}\end{array}
&\begin{array}{c}
\hbox{faithfully}\cr
\hbox{represented}\cr
\hbox{internal group}\end{array}
&\begin{array}{c}
\hbox{faithfully}\cr
\hbox{represented}\cr
\hbox{external group}\end{array}
\cr\hline\hline
0&\C&\{1\}&\{1\}\cr\hline
1&\UU,\UU^*\cong\C^2&\U(2)&\UL(2)\cr\hline
2&
\begin{array}{rl}
\UU\ox\UU^*&\cong\C^4\cr
\UU\and\UU,\UU^*\and\UU^*&\cong\C\cr
\end{array} 
&\begin{array}{c}
\SO(3)\cr
\U(1)\cr
\end{array}
&\begin{array}{c}
\SO^+(1,3)\cr
\U(1)\cr
\end{array}
\cr\hline
3&
\begin{array}{rl}
\UU\ox\UU^*\and \UU^*&\cong\C^2\cr
\UU\and\UU\ox\UU^*&\cong\C^2\cr
\end{array}&
\U(2)&\UL(2)\cr\hline
4&(\UU\and\UU)\ox(\UU\and\UU)^*\cong\C&\{1\}&\{1\}\cr\hline\end{array}\cr\cr
\hbox{\it  $\U(2)$- and $\UL(2)$-properties of the 
Grassmann algebra ${\bf GRASS}$} 
\end{eq}

With the basic fermion field $\psi$ the 
internal hyperisospin $\U(2)$-properties of a reference
Grassmann algebra for $(\UU\pl\UU^*)(0)$ are considered in 
the external Lorentz phase group $\UL(2)$-properties 
of a Grassmann algebra for $(\UU\pl\UU^*)(x)$
\begin{eq}{l}
{\AND}(\psi\pl\psi^*)(x):
{\bf GRASS}(x)\map {\bf GRASS}(0)
\end{eq}with isomorphism between corresponding  vector subspaces
with  an corresponding external and internal representation structure
\begin{eq}{c}\hskip-3mm
\begin{array}{|c|c|c|c|}\hline
  N &\hbox{basic field}
&\begin{array}{c}
\!\!\!\U(2)\!=\!\U(\bl1_2)\o\SU(2)\!\!\!\cr
[y||2T]\end{array}
&\begin{array}{c}
\!\!\!\UL(2)\!=\!\U(\bl1_2)\o\SL(\C^2)\!\!\!\cr
[c||2J_L|2J_R]\end{array}
\cr\hline\hline
0&\id_\C
&[0||0]
&[0||0|0]
\cr\hline
1&\psi(x),~\psi^*(x)
&[-{1\over2}||1],~[+{1\over2}||1]
&[+{1\over2}||1|0],~[-{1\over2}||0|1]
\cr\hline
2&\begin{array}{c}
(\psi\ox\psi^*)(x)\cr
(\psi\and\psi)(x),~(\psi\and\psi)^*(x)\end{array}
&\begin{array}{c}
[0||0]\pl[0||2]\cr
[\mp1||0]\end{array}
&\begin{array}{c}
[0||1|1]\cr
[\pm1||0|0]\end{array}
\cr\hline
3&\begin{array}{c}
(\psi\ox\psi^*\and\psi^*)(x)\cr(\psi\and\psi\ox\psi^*)(x)\end{array}
&\begin{array}{c}
[+{1\over2}||1]\cr
[-{1\over2}||1]\end{array}
&\begin{array}{c}
[-{1\over2}|1||0]\cr[+{1\over2}|0||1]\end{array}
\cr\hline
4&(\psi\and\psi)\ox(\psi\and\psi)^*(x)
&[0||0]
&[0||0|0]
\cr\hline
\end{array}\cr\cr
\hbox{\it quantum numbers of the basic field products}
\end{eq}

A basic field $\psi$, quantized with anticommutators, cannot
imbed the $\U(1)$-properties of 
$\UU\and\UU\cong\C$ with Grassmann degree $N=2$, since
the scalar combination vanishes
\begin{eq}{l}
\psi\and\psi(x): ~~
\psi_A^\al\ep^{ AB}\ep_{\al\be}
\psi_B^\be(x)=
{1\over2}\ep^{AB}\ep_{\al\be}
\acom{\psi_A^\al(x)}{\psi_B^\be(x)}=0
\end{eq}Only the combination leading to an $\SU(2)$-triplet is nontrivial
\begin{eq}{l}
\psi\and\psi(x)\sim
\psi_A^\al\ep^{AB}\rvec\tau_{\al\be}
\psi_B^\be(x),~~\rvec\tau_{\al\be}=\ep_{\al\ga}\rvec\tau_\be^\ga
=\rvec\tau_{\be\al}
\end{eq}

Therewith one has to consider 
four types of nontrivial  fields - two fermionic fields
with odd Grassmann degree $1$ and $3$ and two bosonic 
fields with even Grassmann degree $2$ and $4$.
Only $N=1,2,3$  
characterize  nontrivial symmetric spaces and 
representations of the nonabelian boost manifold (conjugation manifold)
$\UL(2)/\U(2)$
\begin{eq}{c}
\begin{array}{|c|c|c|c|}\hline
  N &{z\over2}&\hbox{basic field}&\hbox{manifold representation}\cr\hline\hline
1&\mp {1\over2}&
\psi(x),~~\psi^*(x)&\UL(2)/\U(2)\cr\hline
2&0&(\psi\ox\psi^*)(x)&\SO^+(1,3)/\{1\}\pl \SO(3)\cr\hline
3&\pm {1\over2}&\begin{array}{c}
(\psi\ox\psi^*\and\psi^*)(x)\cr
(\psi\and\psi\ox\psi^*)(x)
\end{array}&\UL(2)/\U(2)\cr\hline
4&0&(\psi\and\psi)\ox(\psi\and\psi)^*(x)
&\{1\}\cr\hline
\end{array}\cr\cr
\hbox{\it $\UL(2)/\U(2)$-representations by basic field products}
\end{eq}

In addition to the $\D(1)$-grading with the 
natural number Grassmann degree $N\in\N$,
a Grassmann algebra over a selfdual complex space $\UU\pl\UU^*\cong\C^{2n}$
has a $\U(1)$-grading with $z\in\Z_{2n+1}$.
The $\U(1)$-property defines the hypercharge 
 and chirality $\Z_5$-grading with $y,c={z\over2}=0,\pm{1\over2},\pm1$ 
\begin{eq}{l}
{\bf GRASS}={\PL_{z=-2}^2}\UU^{(z)},\left\{\begin{array}{rl}
\UU^{(0)}&=\C\pl[\UU\ox\UU^*]\pl[\UU\and\UU\ox\UU^*\and\UU^*]\cong\C^6\cr 
\UU^{(1)}&=\UU^*\pl[\UU\ox\UU^*\and\UU^*]\cong\C^4\cr 
\UU^{(-1)}&=\UU\pl[\UU\and\UU\ox\UU^*]\cong\C^4\cr 
\UU^{(2)}&=\UU^*\and\UU^*\cong\C\cr
\UU^{(-2)}&=\UU\and\UU\cong\C\cr 
\end{array}\right. 
\end{eq}

A basic theory for the symmetric space $\GL(\C^2)/\U(2)$ has to use
only the  field $\psi$ in analogy to the tetrad $h$ for 
minimal  relativity $\GL(\R^4)/\O(1,3)$.
The standard model is not basic in this sense. But, at least,
the correspondence between the relevant basic field products and the 
effective particle oriented standard fields can be found.

\section{External-Internal Cosets in the Lepton Model}

The `colourless' standard model, i.e. without quark and gluon fields,
pa\-ra\-me\-tri\-zes all nontrivial external-internal or internal-external 
cosets, $G_{\rm ext}/G_{\rm int }$ and 
$G_{\rm int}/G_{\rm ext}$ resp., 
which are possible with the $\UL(2)$ and $\U(2)$-re\-pre\-sen\-ta\-tions
in the Grassmann algebra 
\begin{eq}{c}
\begin{array}{|c||c|c||c|}\hline
&\UL(2)&\U(1)&\SO^+(1,3)\cr\hline\hline
\U(2)&\bl l &\bl H&\x\cr\hline
\U(1)&\bl e & - &\x\cr\hline\hline
\{1\}\pl\SO(3)&\x &\x&\bl A\pl \bl B\cr\hline
\end{array}
\end{eq}One has to consider
the possibilities to
embed into  each other 
the nontrivial external groups $\UL(2)$, $\U(1)$ and $\SO^+(1,3)$
and the nontrivial  internal groups $\U(2)$, $\U(1)$ and $\SO(3)$
- in both directions.

Internal $\U(2)$ can be embedded only in external $\UL(2)$, done by the
lepton isodoublet fields, 
as flat space contribution for the basic fields $\psi,\psi^*$ 
\begin{eq}{l}
\UL(2)/\U(2):~~
\left\{\begin{array}{rll}
\bl l(x):&\UU(x)\map \UU(0),~&\bl l(x)\sim\bl l_A^\al(x)\cr
\bl l^*(x):&\UU^*(x)\map\UU^*(0)
,~&\bl l^*(x)\sim\bl l_\al^{*\dot A}(x)\end{array}\right.
\end{eq}

Internal $\SO(3)$ can be embedded only in external $\SO^+(1,3)$, done
by the gauge fields, corresponding to the basic field $\psi\ox\psi^*$ 
\begin{eq}{rl}
\SO^+(1,3)/\{1\}\pl \SO(3):
&\bl A(x)\pl\bl B(x):(\UU\ox\UU^*)(x)\map(\UU\ox \UU^*)(0)
\cr
~~&\bl A(x)\pl\bl B(x)\sim\bl A_j(x)+\rvec{\bl B}_j(x)\cr
\end{eq}

The embedding of external
$\U(1)$ in internal $\U(1)$ is  trivial  $\U(1)/\U(1)\cong\{1\}$.

Internal $\U(2)$ can imbed only external $\U(1)$, done by the Higgs isodoublet fields
\begin{eq}{l}
\U(2)/\U(1):~~\left\{\begin{array}{rll}
\bl H(x):&(\UU\and\UU)(x)\map\UU(0),~&\bl H(x)\sim\bl H^\al(x)\cr
\bl H^*(x):&(\UU\and\UU)^*(x)\map \UU^*(0),~&\bl H^*(x)\sim\bl H^*_\al(x)
\end{array}\right.
\end{eq}

Internal $\U(1)$ can be embedded only in external $\UL(2)$, done by
the lepton isosinglet fields 
\begin{eq}{l}
\UL(2)/\U(1):~~\left\{\begin{array}{rll}
\bl e(x):&
\UU^*(x)\map(\UU\and\UU)(0),~&\bl e(x)\sim\bl e_{\dot A}(x)\cr
\bl e^*(x):&\UU(x)\map(\UU\and\UU)^*(0)
,~&\bl e^*(x)\sim\bl e^{* A}(x)\end{array}\right.
\end{eq}

The fields in the diagonal, the $2\x2$
lepton fields $\bl l$ (isodoublet Lorentz-doublet)
  and 
 the $4\x4 $ gauge 
 fields $\bl A\pl\bl B$ ($\U(2)$-quartet Lorentz-vector) 
connect spaces with equal dimensions.

The pair $(\bl H,\bl e)$  in the
skew-diagonal with
the $2\x 1$ Higgs fields $\bl H$
(isodoublet Lorentz-scalar)  and the 
$1\x 2$ lepton fields  $\bl e$ (isosinglet Lorentz-doublet) 
come together  as a `doublet property swapping pair'
\begin{eq}{l}
[y||2T]\o[c||2J_L|2J_R]=\left\{
\begin{array}{rcrl}
[1||0]&\o&[-{3\over2}||1|0]&\hbox{for }\bl e^*\cr
[-{1\over2}||1]&\o&[1||0|0]&\hbox{for }\bl H\cr
[{1\over2}||1]&\o&[-{1\over2}||1|0]&\hbox{for }\bl e^*\ox \bl H\cr
\end{array}\right.
\end{eq}The internal $\SU(2)$ for the Higgs Lorentz singlet 
field $\bl H$
corresponds to the external 
$\SU(2)\subnoteq \SL(\C^2)$ for the lepton isosinglet
\begin{eq}{l}
{\U(1)\x\UL(2)\over \U(2)\x\U(1)}\cong\UL(2)/\U(2)
\end{eq}The $\SU(2)$-swapping pair can arise from the isomorphisms
for the tensors of Grassmann degree 3
\begin{eq}{l} 
\chi(x):(\UU\ox\UU^*\and\UU^* )(x)\map (\UU\ox \UU^*\and \UU^*)(0)\cr
\chi(x)=(\psi\ox \psi^*\and\psi^*)(x)\sim
\chi_A^\al(x)=
\psi_ A^\be\rvec\tau_\be^\al ~
\psi^{*\dot C}_\ga\ep_{\dot C\dot D}
\rvec\tau^{\ga\de}\psi^{*\dot D}_\de(x)\cr
\end{eq}as a  particle oriented 
twofold factorization in  the flat spacetime expansion
\begin{eq}{rlll}
(\psi\ox\psi^*\and\psi^*)(x)&=(\bl e^*\ox\bl H)(x)+\dots,&
\chi_A^\al(x)&=\ep_{AB}\bl e^{*B}\bl H^\al(x)+\dots\cr
(\psi\and \psi\ox\psi^*)(x)&=(\bl H^*\ox\bl e)(x)+\dots,&
\chi^{*\dot A}_\al(x)&=\ep^{\dot A\dot B}\bl H^*_\al\bl e_{\dot B}(x)+\dots\cr
\end{eq}

\section{Quark Fields as Grassmann Roots}

The main problem for 
an interpretation of  the standard model 
in the framework of a basic $\GL(\C^2)/\U(2)$
coset structure are the coloured fields,
the quark fields $\bl q,\bl d,\bl u$  and the gluon fields $\bl G$.
The only natural relation of $\U(2\x3)$ to $\U(2)$
seems to arise in the Grassmann algebra ${\bf GRASS}\cong\C^{16}$
over $\UU\pl\UU^*\cong\C^4$ which gives rise to 
two types of faithful $\U(2)$-re\-pre\-sen\-ta\-tions with Grassmann
degree  $ N =1$ and $  N =3$
which may reflect colour singlet and colour triplet properties resp.
 In analogy to the 
representation of $\UU\ox(\UU\and\UU)^*$ with the Higgs-lepton 
two factor product
$\bl e^*\ox\bl H$ the quarks may arise from a parametrization 
with a three factor product
\begin{eq}{l}
\UU\ox\UU^*\and\UU^*:\left\{
\begin{array}{cll}
\U(2)\ox\U(2))\and\U(2)&\cong\U(2)\ox\U(1)&\cong\U(2)\cr
\UL(2)\ox\UL(2))\and\UL(2)&\cong\UL(2)\ox\U(1)&\cong\UL(2)\end{array}\right.
\end{eq}taking care of the 
$\GL(\C)=\D(1)\x\U(1)$-properties given by
the two gradings of the Grassmann algebra.

Originally, the quarks were
introduced  as `cubic root'-re\-pre\-sen\-ta\-tions of the nucleons
with colour $\SU(3)$ as gauge group for the strong interactions.
As seen in the standard model central correlation 
$\I(3)\cong\SU(3)\cap\U(\bl 1_3)$ (section 4),
a colour $\SU(3)$-property with nontrivial triality
 \cite{BIED},
i.e. an $\SU(3)$-re\-pre\-sen\-ta\-tion $[ C_1, C_2]$
with $C_1-C_2\ne 3\Z$, e.g. 
triplets $[1,0]$ or sextets $[2,0]$, not, however, 
octets $[1,1]$ or decuplets $[3,0]$, 
cannot be separated from a third integer hypercharge $\U(1)$-property.

The $\U(3)$-hypercharge-coulour group 
can be
considered to be the  continuous phase generalization of
the discrete cyclotomic root $\exp{{2\pi i\over3}}\in\I(3)$ or, 
general for $\U(N)$ with $\centr\U(N)\cong\I(N)$ 
\begin{eq}{l}
k=1,\dots,  N :\left\{
\begin{array}{ccccl}
\exp{{2\pi ik\over  N }}=&[\exp{{2\pi i\over  N }}]^k&\hbox{with}
&[\exp{{2\pi ik\over  N }}]^ N &=1\cr  
{\AND^k}\U( N )=&\{ {\AND^k} u\mid u\in\U( N )\}&\hbox{with}
&{\AND^ N }[{\AND^k}\U( N )]&\cong\U(1)\end{array}\right.
\end{eq}Any root $\exp{{2\pi ik\over  N }}\in\I( N )$ 
as power of the cyclic root $\exp{{2\pi i\over  N }}$
has its correspondence
in the group which is defined by the   $\U( N )$ representation
${\AND^k}\U( N )$ on a complex ${ N \choose k}$-di\-men\-sio\-nal space
as   $k$-th Grassmann  power of the cyclic defining representation
with $k=1$
\begin{eq}{rl}
\I( N )=\sqrt[ N ]1&=\{\exp{{2\pi ik\over  N }}\mid k=1,\dots,  N \}\cr
\sqrt[ N ]{\U(1)}&=\{{\AND^k}\U( N )\mid k=1,\dots, N \}
\hbox{ with }\U( N )\cong{\U(1)\x\SU( N )\over\I( N )}
\end{eq}Here are as examples the
2nd, 3rd and 6th Grassmann roots of $\U(1)$ 
\begin{eq}{c}
\begin{array}{|c||c|c|}\hline
k&1&2\cr\hline\hline
\sqrt[2]1&\exp{2\pi i\over2}&\exp{2\pi i\over1}\cr\hline
\sqrt[2]{\U(1)}&\U(2)&\U(1)\cr\hline
\end{array}~~~~~~
\begin{array}{|c||c|c|c|}\hline
k&1&2&3\cr\hline\hline
\sqrt[3]1&\exp{2\pi i\over3}&\exp-{2\pi i\over3}&\exp{2\pi i\over1}\cr\hline
\sqrt[3]{\U(1)}&\U(3)&\U(3)&\U(1)\cr\hline
\end{array}\cr
\cr
\begin{array}{|c||c|c|c|c|c|c|}\hline
k&1&2&3&4&5&6\cr\hline\hline
\sqrt[6]1&\exp{2\pi i\over6}&\exp{2\pi i\over3}&\exp{2\pi i\over2}
&\exp-{2\pi i\over3}&\exp-{2\pi i\over6}&\exp{2\pi i\over1}\cr\hline
\sqrt[6]{\U(1)}&\U(6)&{\U(6)\over\I(2)}&{\U(6)\over\I(3)}
&{\U(6)\over\I(2)}&\U(6)&\U(1)\cr\hline
\end{array}\cr
\end{eq}

With Grassmann powers one can define
the $ N $-th Grassmann root of $\U(n)$ for relatively
prime\footnote{\scriptsize
The Grassmann root of $\U(1)$ for any natural number $m=1,2,\dots$
is obtained by using its Sylow decomposition  
$m=p_1^{k_1}\cdots p_r^{k_r}$ in powers of primes 
\begin{eq}{l}
\sqrt[m]{\U(1)}
=\sqrt[p_1^{k_1}]{\sqrt[p_2^{k_2}]{\cdots \sqrt[p_r^{k_r}]{\U(1)}}}
=\U(p_1^{k_1}\x\cdots\x p_r^{k_r})=
{\U(1)\x\SU(p_1^{k_1})\x\cdots\x\SU( p_r^{k_r})\over
\I(p_1^{k_1})\x\cdots\x\I( p_r^{k_r})}
\end{eq}
} naturals $(n, N )$,
e.g. for the standard model 
isospin-colour relevant pair  $(n, N )=(2,3)$
\begin{eq}{rl}
\sqrt[ N ]{\U(n)}&=\{{\AND^k}\U(n\x  N )\mid k=1,\dots, N \}
\hbox{ with }\U(n\x  N )={\U(1)\x\SU(n)\x\SU( N )\over\I(n)\x\I( N )}
\end{eq}The root allows the distribution of the $\U(1)$-phase in
$\U(n)$ on $k\le  N $ factors, e.g. for $(n, N )=(2,3)$
\begin{eq}{c}
\begin{array}{|c||c|c|c|}\hline
k&1&2&3\cr\hline\hline
\sqrt[3]{\sqrt[2]1}&\exp\pm{2\pi i\over6}&\exp\pm{2\pi i\over3}
&\exp{2\pi i\over2}\cr\hline
\sqrt[3]{\U(2)}&\U(2\x 3)&\U(3)&\U(2)\cr\hline
\end{array}
\end{eq}

The quark fields as cubic Grassmann roots 
take care of the basic field products  with Grassmann degree $ N =3$ in 
$\UU\ox\UU^*\and\UU^*$, $\UU\and\UU\ox\UU^*\cong\C^2$.
The quark isodoublet field $\bl q$
parametrizes the $k=1$ member of the internal $\U(2)$-roots
$\sqrt[3]{\U(2)}$ with $\U(2\x3)$-degrees of freedom, 
the two quark isosinglets $\bl d,\bl u$ parametrize   
the $k=2$ member of $\sqrt[3]{\U(2)}$ with
$\U(3)$-degrees of freedom 
\begin{eq}{l}
\sqrt[3]{\psi\ox\psi^*\and\psi^*} 
=\left\{
\begin{array}{lr}
\bl q +\dots,& k=1\cr
\bl q\and\bl q \pl\bl d\and\bl u +\dots ,&k=2 \cr
\bl q\and \bl q\and\bl q \pl\bl q\and\bl d\and\bl u +\dots,&k=3
\end{array}\right.\cr
~~~~~\sim\left\{
\begin{array}{l}
\bl q_A^{\al c} +\dots\cr
 \ep_{c_1c_2c_3}\ep_{\al_2\al_3}
(\bl q_{A_2}^{\al_2 c_2}\ep^{A_2A_3}\bl q_{A_3}^{\al_3 c_3}+
\bl d_{\dot A_2}^{\al_2 c_2}
\ep^{\dot A_2\dot A_3}\bl u_{\dot A_3}^{\al_3 c_3}) +\dots\cr
 \ep_{c_1c_2c_3}\ep_{\al_2\al_3}\bl q_A^{\al c_1}
(\bl q_{A_2}^{\al_2 c_2}\ep^{A_2A_3}\bl q_{A_3}^{\al_3 c_3}+
\bl d_{\dot A_2}^{\al_2 c_2}
\ep^{\dot A_2\dot A_3}\bl u_{\dot A_3}^{\al_3 c_3}) +\dots\cr
\end{array}\right.\cr
\end{eq}which is written with the representations 
\begin{eq}{rl}
\sqrt[3]{
[{1\over2}||1]}=\left\{\begin{array}{rl}
[{1\over6}||1;1,0] ,&k=1\cr
[{1\over3}||0;0,1],&k=2,~~{1\over3}
={1\over6}+{1\over6}={2\over3}-{1\over3}  \cr
[{1\over2}||1;0,0],&k=3\end{array}\right.  \cr
\end{eq}

If, for  an effective linearization of the
basic $\GL(\C^2)/\U(2)$ coset structure as realized with the
Grassmann algebra ${\bf GRASS}\cong\C^{16}$, 
 the basic internal operation group $\U(2)$ is extended 
 by a cubic Grassmann  root to
 $\U(2\x3)$ one has to
provide also for a gauge field for the additional local
$\U(2\x3)/\U(2)\cong\SU(3)/\I(3)$ operations. This is done 
 in the standard model with the
gluon fields $\bl G(x)$. 
\vskip1cm
\centerline{\bf Acknowledgments}
\vskip5mm
I benefitted from discussions with David Finkelstein and Tony Smith,
both at Georgia Tech, Atlanta.

\end{document}